\documentclass[preprint,tightenlines,showpacs,preprintnumbers,amsmath,amssymb,prb]{revtex4}

\begin{document}
\draft
\title{Optical properties of 4 {\AA} single-walled carbon
nanotubes inside the zeolite channels studied from first principles
calculations}

\author{X.P. Yang, H.M. Weng, and Jinming Dong }
\affiliation{Department of Physics and National Laboratory of Solid
State Microstructures, Nanjing University, Nanjing 210093, P. R.
China }

\begin{abstract}

The structural, electronic, and optical properties of 4 {\AA}
single-walled carbon nanotubes (SWNTs) contained inside the
zeolite channels have been studied based upon the
density-functional theory in the local-density approximation
(LDA). Our calculated results indicate that the relaxed
geometrical structures for the smallest SWNTs in the zeolite
channels are much different from those of the ideal isolated
SWNTs, producing a great effect on their physical properties. It
is found that all three kinds of 4 {\AA} SWNTs can possibly exist
inside the Zeolite channels. Especially, as an example, we have
also studied the coupling effect between the ALPO$_{4}$-5 zeolite
and the tube (5,0) inside it, and found that the zeolite has real
effects on the electronic structure and optical properties of the
inside (5,0) tube.

\end{abstract}

\date{16 July 2007}
\maketitle

\section{\label{sec:intro}Introduction}

In the past decade carbon nanotubes (CNTs) [1-5] had been
extensively investigated, which is motivated both by their
electrical and mechanical properties as well as by their potential
applications in future's nanostructured materials. For example,
nanotubes are ideal model systems for studying the transport of
electrons in one dimension. Their unique electronic and mechanical
properties are proved to be a rich source of new fundamental
physics and also make CNTs promising candidates as nanoscale
wires, transistors and sensors.

The single-walled carbon nanotube is the simplest type of carbon nanotubes,
discovered first by Iijima group [6] in 1993, which is composed of rolled up
2D-graphite sheet. The carbon atoms on the SWNT are arranged in a helical
line around its axis. It is well-known that diameter of the SWNTs strongly
affects their physical properties, and it has been proved [7] that the
calculated curvature energies are inversely proportional to the square of
tube radius. The variation of the band gaps of the semiconducting SWNTs with
their radius is found to follow the simple rule based on the zone folding
theory only for larger diameter tubes, but to differ from it for the tubes
with smaller radius. So, it is greatly interested, both theoretically and
experimentally [8-11], to study the physical properties of the SWNTs with
possible smallest diameters.

The smallest SWNTs with 4 {\AA} diameter have been recently produced in the
1 nm-sized channels of the ALPO$_{4}$-5 single crystals (AFI in the zeolite
terminology) [12], which have the following possible chiral geometrical
structures: zigzag (5,0) (diameter, $d $= 3.93 {\AA}), armchair (3,3) ($d $= 4.07
{\AA}) and chiral (4,2) ($d $= 4.14 {\AA}). For such small nanotubes, their
large curvature leads to a hybridization of $\sigma ^{\ast }$and $\pi
^{\ast }$orbitals [13], which has great effects on their electronic
structure. For example, unlike larger diameter SWNTs, which can be either
metallic or semiconducting, depending only on their helicity, now the zigzag
(5,0) tube becomes a metal. Therefore, it is an interesting and also a great
challenge to investigate experimentally and theoretically physical
properties of these smallest SWNTs.

It is known that the AFI is a type of microporous crystal with
one--dimensional channels packed in hexagonal arrays. Its framework consists
of alternate tetrahedra of (ALO$_{4})^{ - }$ and (PO$_{4})^{ + }$. The
AFI single crystals are transparent from the near infrared to the
ultraviolet region. Although the AFI is transparent in the specific
frequency range, it does not mean, however, that the AFI has no any effect
on the electronic and optical properties of the SWNTs inside it. Therefore,
it is interesting to see if the AFI crystal has really significant effects
on the physical properties of the smallest SWNTs inside it.

In this paper, we use the first-principles calculation to study the
structural, electronic and optical properties of the 4 {\AA}-diameter SWNTs
inside the AFI single crystal. The geometrical structures of all isolated
three kinds of the smallest tubes are fully relaxed, and compared with other
theoretical results [14,15]. Especially, as an example, the effects of the
zeolite on the electronic structure and optical absorption of the zigzag
(5,0) tube have been studied.

\section{\label{sec:method} Computational method and details}

The total energy plane-wave pseudopotential method has been used in our
calculations within the framework of local density approximation (LDA), in
which the exchange-correlation energy of the Ceperley and Alder form [16]
was included. The ion-electron interaction is modeled by ultra-soft local
pseudopotentials of the Vanderbilt form [17] for the carbon atoms with
maximum plane wave cut-off energy of 280 eV. The plane-wave pseudopotential
program, CASTEP [18], is used on the selected systems.

The tube bundle in the channels of AFI is modeled by using a supercell
geometry [13], so that the tubes are aligned in a hexagonal array with the
closest distance between the adjacent tubes being 13.726 {\AA}, which is the
same as that between the adjacent axis of the AFI channels, and found to be
a larger enough to prevent the tube-tube interactions. The space group was
$P_{1}$ for the computational models involved in this paper. Firstly, we
optimized the supercell of the SWNTs employing BFGS geometry optimization
scheme [19,20], and the Monkhorst-Pack scheme [21] with a distance of
0.04/{\AA} between the sampling points in the reciprocal space. The BFGS
scheme allows us to specify constraints on the lattice constants and angles
before calculation and optimize the cell during the course of calculation.
After the final self-consistency cycle, the remaining forces on all atoms
were less than 0.03 eV/{\AA}, and the remaining stress was less than 0.05
GPa. Secondly, the calculations on band structure and optical properties are
carried out on the relaxed tube bundles. The calculation of energy band
structure in reciprocal space are performed over 21 $k$ points along the tube
direction, and in the calculation of the polarized absorption spectra the
Monkhorst-Pack scheme with a distance of 0.02/{\AA} between points is used
for the sampling of reciprocal space. In addition, we also carry out the
same calculations on three kinds of tube bundles by using the
Troullier-Martin norm-conserving nonlocal pseudopotentials [22] in the
Kleinman-Bylander form [23] for comparison, and found that obtained results
by using the two different kinds of methods are equivalent to each other.

Also, in order to understand the effects of the AFI crystal on tubes inside
its channels, we modeled the combined structure of the AFI crystal with
zigzag (5,0) tubes in its channels as shown in Fig. 1. The same lattice
constant and symmetry group as those without the AFI crystal are taken for
this combined structure, in which the geometrical structures of the zigzag
(5,0) tubes had been first relaxed. Here, the reason of selecting zigzag
(5,0) tube from the three kinds of SWNTs is because its lattice constant
along the tube direction (c = 4.23 {\AA}) is about half that of the AFI
crystal (c = 8.484 {\AA}), which makes the AFI structure to match more
easily with the zigzag (5,0) tubes in its channels than with other kinds of
SWNTs. The total atom numbers in one supercell of the combined structure is
112 (72 for the AFI crystal and 40 for zigzag tube), and so only ultra-soft
pseudopotentials can be used in our calculation for this combined structure,
taking into account the computational time and cost. The combined structure
has been entirely relaxed by the same BFGS geometry optimization scheme.

\section{\label{sec:result} Results and discussions }

\subsection{Tube bundles (5,0), (3,3) and (4,2)}

The obtained structure parameters for the fully relaxed pure tube
bundle are shown in Table 1. The radius of each tube is slightly
larger than that from an ideal rolling of a graphite sheet, while
the lattice constant along the tube axis is smaller than that of a
rolled-up graphite. We also calculated the bond lengths of these
tubes, and find that the average bond length along the tube axis
is shorter than those of the strained bonds in the direction of
the circumference. In addition, we find the cohesive energies per
carbon atom for three kinds of nanotubes are similar to each
other, with only a slight difference of $E_{(4,2)} < E_{(3,3)} <
E_{(5,0)} $, while the radius of these tubes are in the reverse
order, $r_{(4,2)} > r_{(3,3)} > r_{(5,0)} $. Our results of the
geometry optimization are similar to those obtained by others
[14,15]. The calculated electronic band structures for the three
SWNTs are shown in Fig. 3, in which the zero of energy is set at
the top of the valence band. High-symmetry points in the brillouin
zone (BZ) include: $\Gamma $ = (0, 0, 0) and ${\rm X}$ = (0, 0,
0.5). We also find the zigzag (5,0) tubes are more sensitive to
the structure optimization than the armchair (3,3) and the chiral
(4,2) tubes in our calculations. It is seen from Fig. 3 that the
zigzag (5,0) tube now becomes metallic due to its heavy curvature
effects, which is also in consistent with other first-principles
calculations [14,15]. In addition, the semiconducting (4,2) tube
now has a small indirect energy gap of 0.25 eV rather than the
direct one of 2 eV obtained from the zone-folding. The armchair
(3,3) tube is still metallic, as expected.

Our calculated results for the absorption spectra polarized parallel to the
tube direction are given in Fig. 4 and Table 2. A Gaussian smearing
of 0.1 eV is used in the calculation of the optical spectra. It is seen from
them that there are three strong peaks at 1.48 eV, 2.88 eV, 3.43 eV and a
weaker peak near 2.5 eV for the zigzag (5,0) tube; a sharp peak at 3.3 eV,
and a weaker peak at 2.87 eV for the armchair (3,3) tube; and a strong peak
at 2.24 eV, and two weaker peaks at 3.04 eV and 3.8 eV for the chiral (4,2)
tube. Comparing with the experimental data [24], we can identify the feature
A (1.37 eV) is contributed from the peak at 1.48 eV for (5,0) tube, and the
feature B (2.1 eV) from the peak at 2.24 eV for (4,2) tube and also the peak
at 2.5 eV for (5,0) tube. Finally, the feature C (3.1 eV) is coming from all
the three kinds of tubes.

For comparison, the same kind of calculation is performed on the three tubes
by using the Troullier-Martin norm-conserving non-local pseudopotentials in
the Kleinman-Bylander form. It is found that the calculated results are the
same as those listed above, showing the suitability of the ultra-soft local
pseudopotentials used in our calculations. From the comparison between our
calculated results and the experimental observations, it is concluded that
all the three kinds of 4 {\AA} tubes can possibly exist in the channels of
the AFI crystal. In addition, we find our results for the absorption peaks
are roughly the same as those obtained by others except for our peak
position are a little blue-shifted compared with those in ref. [14,15]. In
order to identify correctness of our calculated results, the full potential
linearized augmented plane-wave (FP-LAPW) [25] calculation has been
performed on the zigzag (5,0) tubes, and obtained result is shown in Fig. 4d
and Table 2. A Gaussian smearing of 0.05 eV is used in the FP-LAPW
calculation of the optical spectra. Comparing Fig. 4d with 4a, we can find
that our results on (5,0) tube obtained by the ultra-soft pseudopotentials
is much closer to those by the FP-LAPW method, especially for peak
positions.

\subsection{The combined structure of the AFI crystal with the relaxed
zigzag (5,0) tube inside the channels}

In order to see the AFI crystal effect, here, we also carry out the
first-principles calculations on entirely relaxed combined structure of the
AFI crystal with the relaxed zigzag (5,0) tubes inside the channels. Their
structure parameters are shown in Table 3. The calculated
energy band structure and optical absorption spectra polarized parallel to
the channel direction are shown in Fig. 5 and 6,
respectively, in which we used a Gaussian smearing of 0.1 eV for optical
spectra.

Comparing Fig. 5 with Fig. 3a, we can find the energy bands close
to the Fermi level come still from the (5,0) tube since AFI
crystal has a larger energy gap. However, due to the AFI crystal,
the both conduction and valence bands now become closer to the
Fermi surface. It is also obvious that the energy dispersion along
$\Gamma $-${\rm X}$ direction becomes much less than before,
meaning the electron wave functions less extended. Especially, the
bands at X point have been much affected in contrast to at the
$\Gamma $ point due to the doubling of the period imposed by the
zeolite, which causes the tube-like bands close to Fermi level in
Fig. 5 to be folded at the X boundary. In addition, a lot of
energy bands coming from AFI exist now in the valence bands lower
than -2.7 eV for the combined structure, and in the energy region
of conduction band from 2 to 4 eV some new conduction bands are
found, which are energy bands of the pure (5,0) tube moved down
from those higher than 4 eV. These effects, said above, on the
electronic structure of the pure (5,0) tube will be shown in the
future's relevant experiments. From Fig. 6, it is seen that there
exists a big change for the optical spectra in an energy region
from 7.0 to 12.5 eV, in which appear several high and sharp peaks.
The inset of Fig. 6 shows the allowed dipole transitions in the
energy region lower than 4.0 eV, in which only a slightly change
has been found for the spectra due to effect of the AFI crystal,
as expected. It is seen from the inset, the first peak is
blue-shifted to the relative higher frequency, and the other peaks
are oppositely red-shifted to the lower frequencies. So, the AFI
crystal has real effects on the electronic structure and optical
absorption spectra of the (5,0) tube bundle in it.

In ref. 14, the effective medium theory is used to calculate the effective
dielectric constant of the zeolite-SWNT composite by taking into account the
zeolite effects. They thought that the van der Waals interaction between
nanotube and zeolite is very weak, as well as the influence of the zeolite
on the electronic structure of the nanotube. Therefore, they in fact
neglected this interaction, and treated the zeolite-nanotube composite as a
homogeneous material with an effective dielectric constant. However, our
work really takes into account the interaction between the AFI and (5,0)
tubes by the first-principles calculation, and we find that above
approximation is justified in the long-wavelength limit, and in general, the
AFI has real effect on the absorption spectra.

In our calculation, some factors are not taken into account, for
example, the impurities lying between carbon nanotubes and the AFI
crystal generated experimentally in the course of the pyrolysis of
tripropylamine (TPA) molecules in the AFI channels, which is an
interesting problem, and needs more experimental and theoretical
investigations in future.\\

\small{Acknowledgment: This work was supported by the Natural
Science Foundation of China under Grant No. 10074026. The authors
acknowledge also support from a Grant for State Key Program of
China through Grant No. 1998061407.}

\newpage
\begin{center}
\textbf{Table Captions}
\end{center}

\textbf{Table 1. }Parameters of three kinds of SWNTs in a 13.726
{\AA}$\times $13.726 {\AA}$\times $C$_{l}$ {\AA} hexagonal
supercell. The parameters (a, b, c, $\alpha $, $\beta $, $\gamma
$) are defined as in Fig. 2.

\begin{center}
\begin{table}[htbp]
\begin{tabular}
{|p{32pt}|p{48pt}|p{82pt}|p{30pt}|p{30pt}|p{30pt}|p{30pt}|p{32pt}|p{32pt}|}
\hline Tube& diameter \par (relaxed)& lattice constant C$_{l}$
\par (relaxed)& a& b& c& $\alpha $& $\beta $&
$\gamma $ \\
\hline
(5,0)&
4.03&
4.23&
1.397&
1.440&
&
120.0&
110.9&
 \\
\hline
(3,3)&
4.17&
2.445&
1.420&
1.429&
&
116.1&
118.9&
 \\
\hline
(4,2)&
4.24&
11.204&
1.426&
1.408&
1.432&
113.8&
119.0&
118.7 \\
\hline
\end{tabular}
\label{tab1}
\end{table}
\end{center}

\textbf{Table 2}\textbf{.} The calculated parallel polarized absorption
spectra of the tube bundles. For comparison, also listed are the
experimental results.

\begin{center}
\begin{table}[htbp]
\begin{tabular}
{|p{63pt}|p{57pt}|p{57pt}|p{57pt}|p{66pt}|} \hline Experiment \par
(Ref. 24)& \multicolumn{3}{|p{173pt}|}{Ultra-soft pseudopotentials
} &
FP-LAPW \par (5,0) \\
\cline{2-4}
 &
(5,0)& (3,3)& (4,2)&
\\
\hline (S) 1.2& & & &
 \\
\hline
(A) 1.37&
1.48&
&
&
1.43 \\
\hline
(B) 2.1&
2.5&
&
2.24&
2.52 \\
\hline
(C) 3.1&
2.88, 3.43&
2.87, 3.3&
3.04, 3.8&
2.98, 3.33 \\
\hline
\end{tabular}
\label{tab2}
\end{table}
\end{center}

\newpage
\textbf{Table 3. }Parameters of the combined structure of AFI crystal with
the zigzag (5,0) tube inside the channels. Also listed in it are the
calculated parallel polarized absorption peaks of the tube bundles in 0-4 eV
region.

\begin{center}
\begin{table}[htbp]
\begin{tabular}
{|p{138pt}|p{138pt}|p{138pt}|}
\hline
(5,0) tube&
non-relaxed&
relaxed  \\
\hline
lattice constant C$_{l}$&
8.484&
8.494 \\
\hline
diameter&
4.03&
4.02 \\
\hline a, b, $\alpha $, $\beta $ & 1.397, 1.440, 120.0, 110.9&
1.401,1.438, 120.2, 110.6 \\
\hline
Absorption peaks&
&
1.55, 2.38, 2.77, 3.33 \\
\hline
\end{tabular}
\label{tab3}
\end{table}
\end{center}

\newpage
\begin{center}
\textbf{Figure Captions}
\end{center}

\textbf{FIG. 1. }The model of the AFI crystal structure with the
zigzag (5,0) tubes inside the channels, viewed along the [001]
direction.\\

\textbf{FIG. 2.} Geometry parameters of the 4 {\AA}-diameter
nanotubes.\\

\textbf{FIG. 3. }Calculated energy band structure for three kinds of
nanotubes. (a) zigzag (5,0), (b) armchair (3,3), and (c) chiral
(4,2). The Fermi level is set at zero.\\

\textbf{FIG. 4. }Calculated absorption spectra polarized parallel to
the tube direction for (a) zigzag (5,0), (b) armchair (3,3), and (c)
chiral (4,2) tubes. (d) The FP\textbf{-}LAPW all-electron
calculation results for the zigzag (5,0) tube.\\

\textbf{FIG. 5. }Calculated energy band structure for the entirely
relaxed combined structure of the AFI crystal with the relaxed
zigzag (5,0) tubes inside the channels. The Fermi level is set at
zero.\\

\textbf{FIG. 6. }Calculated absorption spectra polarized parallel to
the channel direction for the relaxed pure zigzag (5,0) (circle),
and the entirely relaxed combined structure of the AFI crystal with
the relaxed zigzag (5,0) tubes inside its channels (star). Inset is
the absorption spectra in the energy region lower than 4.0 eV.

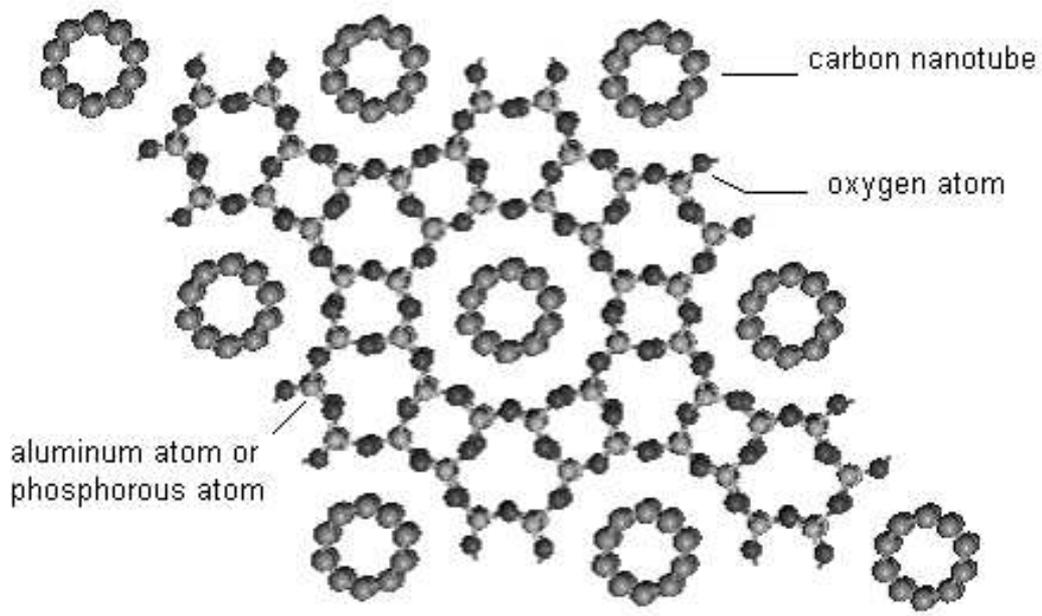
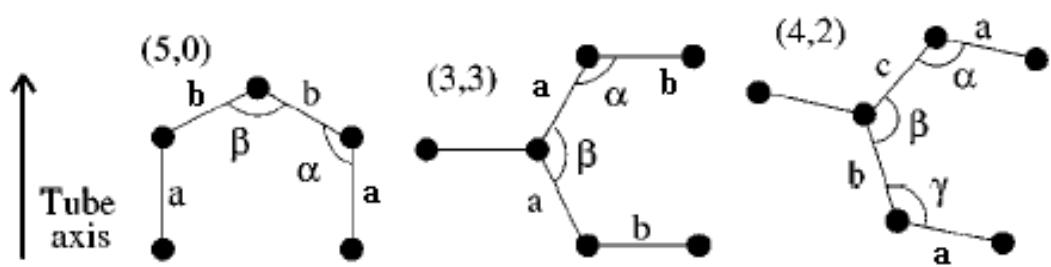
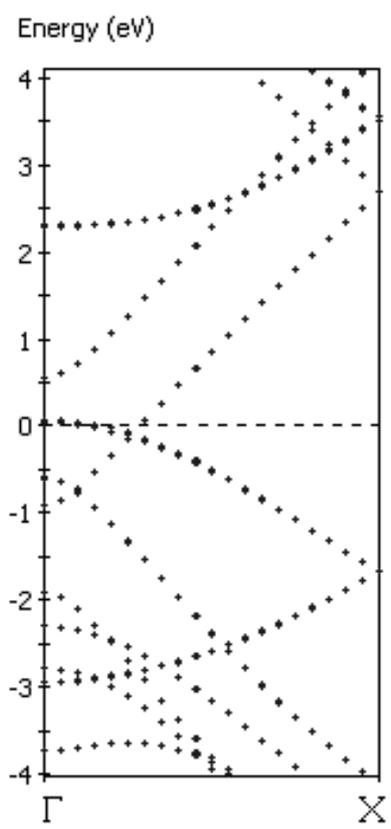
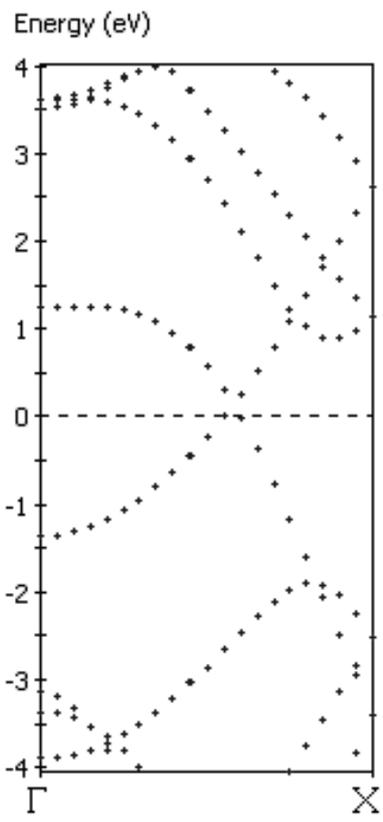
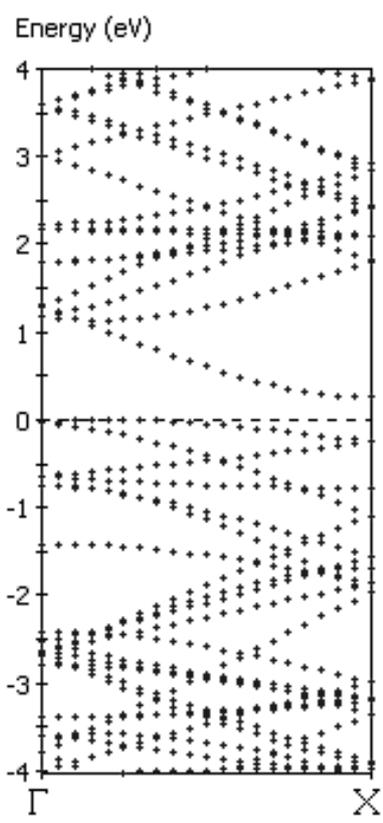
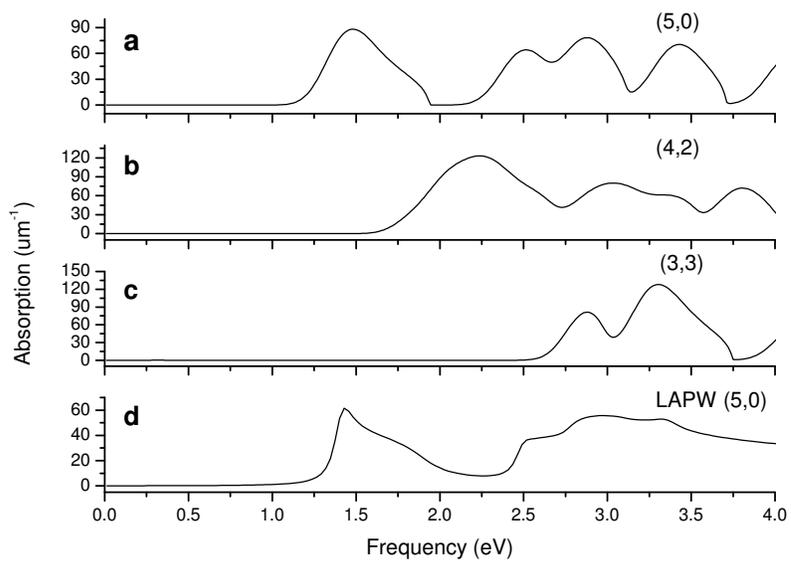
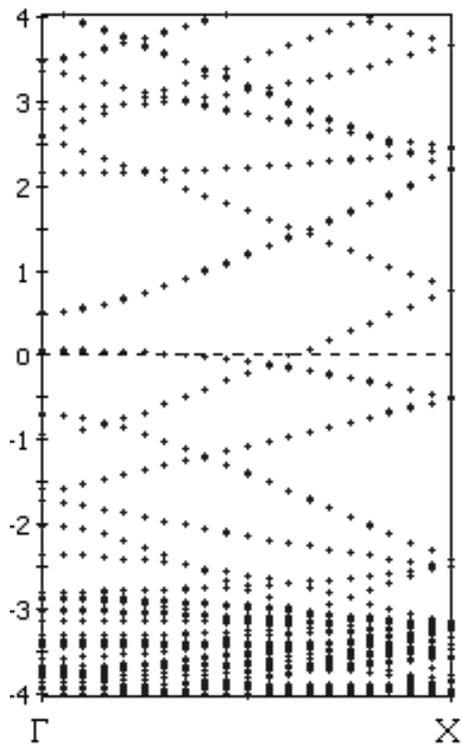
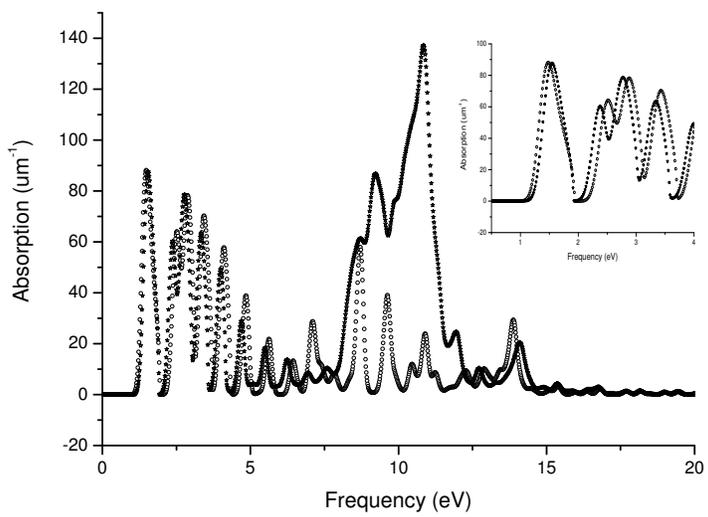
\end{document}